\documentclass{elsart3}

\usepackage[dvips]{graphicx}

\begin{document}


\begin{frontmatter}

\title{ Beam test results of the irradiated Silicon Drift Detector for ALICE }

\author[rez]{S.~Kushpil\thanksref{CA}},
\author[tor]{E.~Crescio},
\author[tor]{P.~Giubellino},
\author[tor]{M.~Idzik},
\author[stb]{A.~Kolozhvari},
\author[rez]{V.~Kushpil},
\author[tor]{M.I.~Martinez},
\author[tor]{G.~Mazza},
\author[rom]{A.~Mazzoni},
\author[rom]{F.~Meddi},
\author[cer]{D.~Nouais},
\author[pra]{V.~Petr$\acute{a}\check{c}$ek},
\author[itc]{C.~Piemonte},
\author[tri]{A.~Rashevsky},
\author[tor]{L.~Riccati},
\author[tor]{A.~Rivetti},
\author[tor]{F.~Tosello},
\author[tri]{A.~Vacchi},
\author[tor]{R.~Wheadon}.

\address[rez]{NPI ASCR $\check{R}e\check{z}$, Czech Republic}
\address[tor]{INFN Sezione di Torino, Italy}
\address[stb]{St.Petersburg University, Russia}
\address[tri]{INFN Sezione di Trieste, Italy}
\address[rom]{INFN Sezione di Roma, Italy}
\address[cer]{CERN, Switzerland}
\address[pra]{Czech Technical University, Prague, Czech Republic}
\address[itc]{ITC-irst, Italy}
\center For the ALICE Collaboration
\thanks[CA]{Corresponding author. E-mail: skushpil@ujf.cas.cz}

\begin{abstract}

The Silicon Drift Detectors will equip two of the six cylindrical
layers of high precision position sensitive detectors in the ITS of the
ALICE experiment at LHC.
In this paper we report the beam test results of a SDD irradiated 
with 1~GeV electrons.
The aim of this test was to verify the radiation tolerance of the device
under an electron fluence equivalent to twice particle fluence
expected during 10 years of ALICE operation.

\end{abstract}

\end{frontmatter}

\section{Introduction}
The Inner Tracking System (ITS) is the central detector of ALICE 
\cite{TDR-ITS,ITS-FLAVIO}. Its basic
functions are the secondary vertex reconstruction of hyperon and charm
decays, the particle identification, the tracking of low-momentum
particles and the improvement of the momentum resolution.
The Silicon Drift Detectors (SDDs) will equip the third
and the fourth layers of the ITS. 
They are very high-resolution non ambiguous two dimensional readout sensors 
adapted to high track density
experiments with low rate because of their relatively slow readout.
Moreover, the operational mode allows a radical reduction in the number
of readout channels.
The ALICE SDDs have to provide a spatial precision of about $30\,\mathrm{\mu m}$ for both 
coordinates.
 Performance of different SDD prototypes has been studied with particle
 beams since 1997 \cite{BTA,BT,NIM05}. In this paper we present the results 
 obtained for detector irradiated by 1GeV electron beam.  
 \section{ Description of the detector }
\label{sec:sddsystem}
  The ALICE SDD final prototypes\cite{SASHAELBA} were produced by Canberra
Semiconductors on $300\,\mathrm{\mu m}$ thick 5'' thick NTD wafers with 
a resistivity of 3~k$\Omega\cdot$cm.
Their active  area is $7.02 \times 7.53\,\mathrm{cm^2}$, i.e.83\% of total area. 
  The active area is split into two adjacent 35 mm long drift regions, 
each equipped with 256 collecting anodes ($294\,\mathrm{\mu m}$ pitch),
with built-in voltage dividers for the drift and the guard regions.
  Design of the cathode strips prevents any punch-through which 
would deteriorate the voltage divider linearity.
  Due to the strong temperature variation of detector's drift
velocity($v\propto T^{-2.4}$), the monitoring of this quantity is
performed by means of three rows of 33 implanted point-like MOS charge
injectors for each drift region \cite{MOSinj1a,MOSinj1b}. 
During SDD operation the hole component of the leakage current is collected 
by the drift cathodes and enters the integrated divider. 
This affects the linearity of the potential distribution on the cathodes themselves 
and, therefore, the position measurement obtained from the drift time. 
Thus it is
critical to monitor such changes in order to be able to
reconstruct potential on the detector at any given time of the experiment .
 This is the purpose of the MOS injectors.
The SDD front-end electronics is based on two 64 channel ASICs named
PASCAL\cite{PASCAL} and AMBRA\cite{AMBRA}.
Four pairs of chips per hybrid are needed to read out 
one half of the SDD. Full description of the electronics 
is given in the paper \cite{ELECTR}. \\  
\indent 
Important steps toward the mass production of the detectors is 
evaluation of their radiation hardness. 
For this study the SDD was irradiated using 
1 GeV electron beam at the LINAC of the Synchrotron in Trieste. 
To reproduce the ALICE radiation environment, the electron fluence 
must be 10 times the pion fluence and 20 times the neutron fluence according 
to the Non-Ionizing Energy Loss (NIEL) hypothesis \cite{NIEL}. For this study 
the electron fluence accumulated by the SDD
is equivalent to the total particle fluence expected during 20 years 
of the ALICE operation and corresponds to an absorbed dose in silicon 
of about 500$\,\mathrm{krad}$. \\ 
\indent
     The laboratory measurements \cite{ELECTPERF} of the anode current and 
     the voltage distribution on the integrated divider as well as 
     and the operation of the MOS injectors demonstrate
that the SDD is sufficiently radiation resistant for the full operation lifetime
of the ALICE experiment. Still, it was necessary to verify these expectations
with a beam test. Within 2002 and 2003 years, the same detector was tested twice
 (before and after its irradiation with electrons) using 
 CERN SPS $\pi^{-}$ beam with $p=100\,\mathrm{GeV/c}$.
  The detector under test was placed on the beam line. A telescope, 
made up of five pairs of single sided silicon strip detectors with
 a strip pitch of $50\,\mathrm{\mu m}$, was used to reconstruct the tracks of passing 
 particles. Precision in the determination of the particle impact point in the 
 SDD plane was $5\,\mathrm{\mu m}$.
 Since the size of the beam spot and the area
covered by the microstrip detectors were smaller than the SDD sensitive area,
the SDD was mounted on a mobile support.  Its position was remotely
controlled and measured with a precision of about $30\,\mathrm{\mu m}$.   
It should be noted that during June 2002 beam test only the central 
anode region of the SDD was studied, and in this case 32-channel PASCAL prototype
was used. To study the irradiated SDD in August 2003 we used 64-channel PASCAL 
to readout full anode array. 
\section{Beam test results}
\label{sec:bt}
\subsection{Cluster Size}

\begin{figure}[h]
  \begin{center}
    \includegraphics[width=7.5cm]{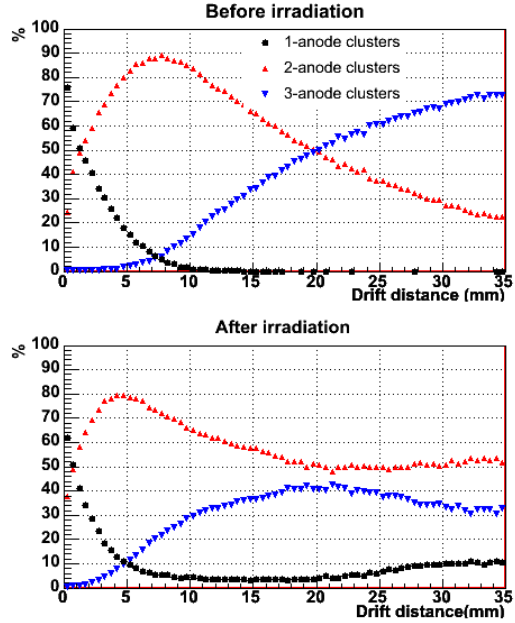}
    \caption{Percentage of the events in which a cluster 
    is collected by one, 
    two or three anodes as a function of the drift distance 
    before and after irradiation.}
    \label{fig:ClSize}
  \end{center}
\end{figure}

The electron cloud generated by an ionizing particle in the SDD undergoes 
a diffusion while drifting to the collection anodes. After the digitization 
of the anode signals, the cloud is represented by a two-dimensional set of
amplitude values, called a "cluster". 
We compared cluster size in the non-irradiated and irradiated detector. 
Fig.~\ref{fig:ClSize} shows the relative amounts of 
clusters collected by one, two and three anodes 
as a function of the drift time. At a short drift distance 
the number of multi-anode clusters increases after irradiation due to 
increased diffusion coefficient. For a large drift distance a presence 
of one-anode clusters can be observed for irradiated detector 
because of a threshold cut
and decrease of the signal amplitude.
\subsection{Charge}
\begin{figure}[h]
  \begin{center}
    \includegraphics[width=7.5cm]{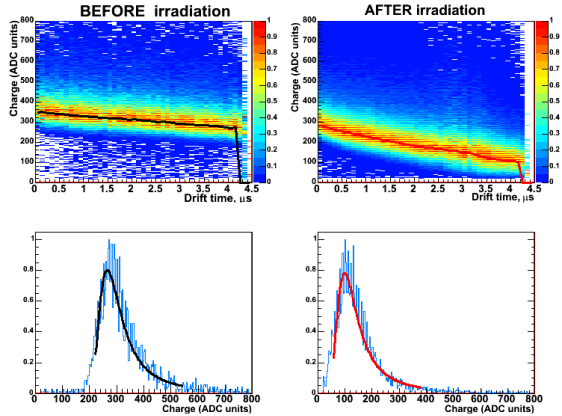}
    \caption{The registered charge as a function of the drift time (top). 
    The example of charge distribution and its fit by the Landau function 
    at drift time of
    4.2$\,\mathrm{\mu s}\, $(bottom). }
    \label{fig:Charge}
  \end{center}
\end{figure}

Fig.\ref{fig:Charge} shows changes in the charge collection in the SDD
before and after irradiation. The collected charge decreases as 
a function of the drift distance. 

A charge collection inefficiency before irradiation was already observed 
in this detector on the test bench in the laboratory. The most probable 
reason is the presence of electron trapping centers 
in the silicon bulk, occasionally introduced in that 
particular wafer during detector fabrication. 
After irradiation a rapidity of charge loss increases by three times due to 
the increased electron trapping. The  comparison of the most probable 
values of the registered charge shows that after irradiation 
the charge collection drops by 60\% at the maximum drift distance.
\subsection{Dopant inhomogeneity} 
Even though the ALICE SDDs are produced on NTD wafers,
which should  have a particularly uniform dopant concentration,
the observed
inhomogeneity characteristic effects  deteriorate significantly 
the spatial resolution of the detectors \cite{ELBADCF,NIM05}.
Inhomogeneity of the dopant concentration alters the uniformity of
the main drift field and, thus, creates systematic deviations 
in the measurement of coordinates of the registered particle.

\begin{figure}[htp]
  \begin{center}
    \includegraphics[width=7.5cm]{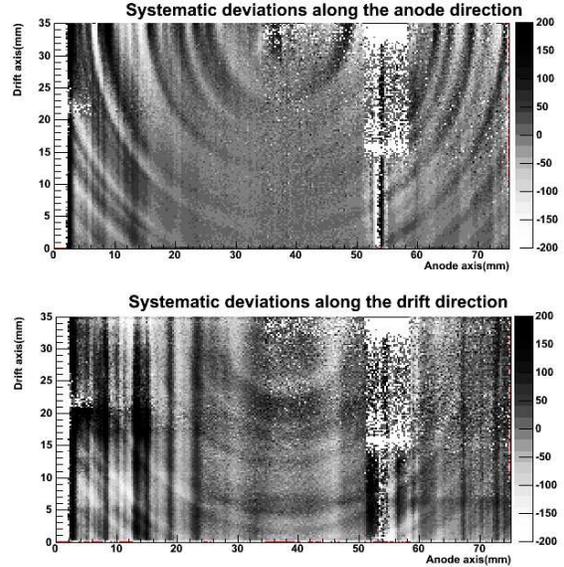}
    \caption{The residuals (grey scale, $\mu m$) of the anode (top) and of 
    the drift (bottom)
    coordinates as a function of the anode coordinate and the drift distance
    for the irradiated SDD.}

    \label{fig:Map}
  \end{center}
\end{figure}

The differences between coordinates of a particle impact point measured by
the SDD and by the microstrip telescope (residuals) are presented in 
Fig.\ref{fig:Map} for the irradiated SDD. They are plotted as functions of 
the anode coordinate and the drift distance. The grey scale represents
magnitude of residuals for the anode  coordinate (top plot) and 
the drift coordinate (bottom plot).
The empty areas correspond to non-working channels or missing experimental data.
Deviation of a few tens of $\mathrm{\mu m}$ in average and with maximum values up
to $200\,\mathrm{\mu m}$ are observed and must be corrected to reach 
the  required spatial resolution  of $30\,\mathrm{\mu m}$.
Recently custom ingots have shown much lower doping fluctuation.
  The circular structures centered in the middle of the wafer clearly visible
in this plot can be attributed to the characteristic radial dependence of the
dopant concentration fluctuations \cite{NIM05,ELBADCF,INCL}. \\

\begin{figure}[htp]
  \begin{center}
    \includegraphics[width=7.5cm]{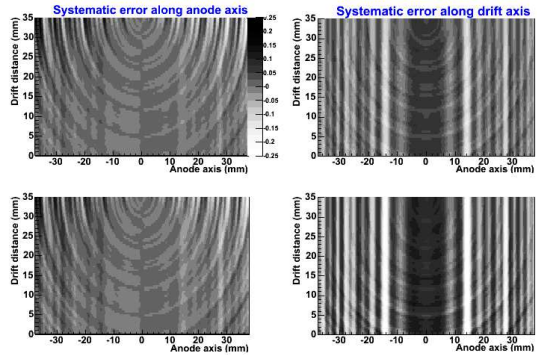}
    \caption{Simulated maps of the systematic deviations before (top) 
    and after (bottom) irradiation.}
    \label{fig:SimMap}
  \end{center}
\end{figure}
In addition to radial structures, the maps present also a deviation 
pattern in a form of vertical lines. Since the effect is similar for all 
electrons collected by a certain anode and looks correlated with the 
intersection of the circular structure by the anode line, we can 
conclude that the local field and its fluctuations in the collection 
region is at the origin of this effect. We can also clearly observe 
that, after irradiation, the magnitude of this linear pattern has increased.
In order to understand whether this evolution of the position correction map 
is easily predictable, a charge transport simulation was performed 
(Fig.\ref{fig:SimMap}), 
taking into account a realistic three-dimensional electrostatic field 
model in the detector. This field was generated by superimposing a 
potential fluctuation map to the solution of the Poisson equation 
assuming a homogeneous silicon bulk. To reproduce qualitatively the 
experimental fluctuation map, the superposition of four radial waves 
with different wavelengths was used.
After irradiation, the difference of potential between adjacent cathodes 
is not anymore constant but assumes a linear evolution, responsible for 
a linear dependence of the electrostatic drift field as a function of 
the drift distance. The drift field is weaker close to the anodes and 
stronger for the maximum drift distance. In order to optionally 
reproduce this effect, a parabolic component can be added to the 
potential in the simulation. The transport calculation of the electrons 
in the silicon bulk takes into account the electrostatic field deriving 
from the previously described potential. The trajectory of the electrons 
was calculated from every node of a grid covering the half SDD surface, 
to the collection anodes.
Assuming a linear trajectory and a constant drift velocity, the initial 
position of the electron can be estimated from its arrival time and 
anode axis coordinate. The two coordinates of the difference of the 
predicted and the actual positions as a function 
of the initial position are plotted in Fig.\ref{fig:SimMap}. 
Two cases are shown: before and after 
irradiation.
The vertical deviation pattern can effectively be observed and its 
magnitude increased when the parabolic potential is added. As a 
conclusion, we can say that the irradiation has only an indirect effect 
on the deviation map through its influence on the voltage divider but no 
significant effect on the  bulk material properties.
\subsection { Spatial resolution}
\begin{figure}[!h] 
 + \begin{center}
    \includegraphics[width=7.5cm]{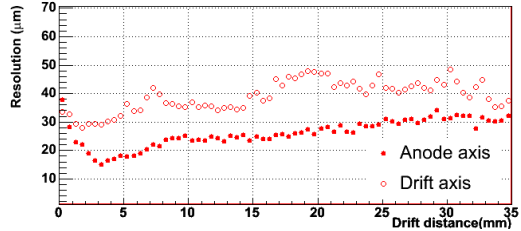}
    \caption{Spatial resolution along the drift and 
    the anode direction as a function of the drift distance. 
    The values were calculated for entire half-size of the irradiated SDD. }
    \label{fig:ResolFull}
  \end{center}
\end{figure}

The detector spatial resolution is defined as the $\mathrm{r.m.s.}$ of the difference 
between the position measured by the SDD and the impact point coordinate reconstructed
with the microstrip telescope.
Fig.\ref{fig:ResolFull} shows the resolution along the anode and the
drift time directions obtained after correction of the systematic
deviation for one half of the irradiated SDD.
The resolution along the anode direction has 
values better than $30\,\mathrm{\mu m}$ 
over more than 70\%  of the whole drift
path and the best value reachs $15\,\mathrm{\mu m}$ at $3\,\mathrm{mm}$ from
the anodes. 
  The deterioration of the resolution at a small drift distance is due to
the a small size of the electron cloud collected on the anodes.
The resolution along the drift direction has a value increasing
from $30\,\mathrm{\mu m}$ to $48\,\mathrm{\mu m}$. 

\begin{figure}[!h]  
  \begin{center}
    \includegraphics[width=7.5cm]{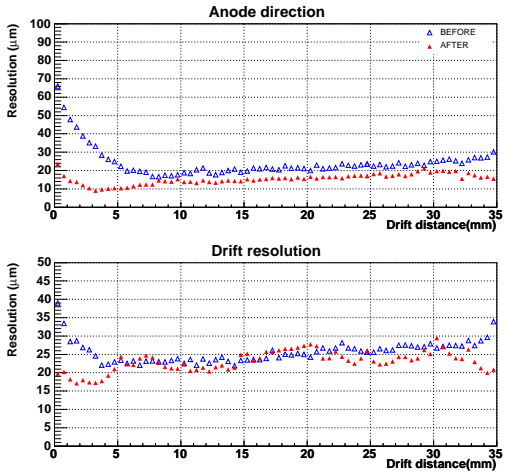}
    \caption{Comparison between the resolution obtained in the narrow 
             central anode region for non-irradiated and irradiated SDD.}
    \label{fig:ResolCenter}
  \end{center}
\end{figure}
For narrow central region  of the SDD anodes it is possible to compare 
the spatial resolution before and after irradiation (Fig.~\ref{fig:ResolCenter}).
One can observe that after irradiation in the vicinity of the anodes, the value 
of the resolution along both direction becomes better.
This behaviour is due to
decreasing fraction of the narrow clusters after irradiation.
For longer drift distances, the values of the resolution are very similar 
to those for non-irradiated detector. Taking into account that the SDD
was irradiated with dose equivalent to 20 years of the ALICE operation, the 
resolution remains within specifications of technical design for the 
ALICE ITS.  Even with the very strong effect 
of the dopant inhomogeneity which increases with irradiation,
it is demonstrated that the systematic deviations in coordinate 
measurements can be corrected and a satisfactory resolution
can be achieved along both anodic and drift directions.
\section{Conclusion}
Extensive study of the performance of a silicon drift detector irradiated 
with dose equivalent to 20 years of the ALICE operation was carried out using
 a 64-channel PASCAL front-end chip. The results show than 
 in spite of increased charge loss the values of the spatial resolution 
  fully satisfy the ALICE technical design requirements, once the correction 
  of the systematic errors is performed. 
 The detector was found to be sufficiently radiation hard 
 for the ALICE experiment.
\vskip 24pt
\noindent{\bf Acknowledgements.\newline\vskip 2pt}

  This work was supported by the grant of the Ministry of education
    of the Czech Republic 1P04LA211 and by the Institutional Research Plan
    AV0Z10480505.

\end{document}